\documentclass[aps,prb,groupedaddress,floatfix]{revtex4}

\usepackage{graphicx}
\usepackage{subfigure}
\usepackage{epsfig}
\usepackage{amsmath}
\usepackage{bbm}

\def\be{\begin{equation}}
\def\ee{\end{equation}}
\def\bea{\begin{eqnarray}}
\def\eea{\end{eqnarray}}

\begin{document}


\title{Quantum Hall effect in carbon nanotubes and curved graphene strips}


\author{E. Perfetto$^{1,3}$, J. Gonz{\'a}lez$^1$, F. Guinea$^2$,
      S. Bellucci$^3$ and P. Onorato$^{3,4}$}
\affiliation{$^1$ Instituto de Estructura de la Materia.
        Consejo Superior de Investigaciones Cient{\'\i}ficas.
        Serrano 123, 28006 Madrid. Spain.  \\
             $^2$Instituto de Ciencia de Materiales.
        Consejo Superior de Investigaciones Cient{\'\i}ficas.
        Cantoblanco. 28049 Madrid. Spain.   \\
             $^3$INFN, Laboratori Nazionali di Frascati,
              P.O. Box 13, 00044 Frascati, Italy. \\
             $^4$Department of Physics "A. Volta", University of Pavia,
          Via Bassi 6, I-27100 Pavia, Italy.}

\date{\today}

\begin{abstract}
We develop a long wavelength approximation in order to describe the
low-energy states of carbon nanotubes in a transverse magnetic field.
We show that, in the limit where the square of the magnetic length
$l = \sqrt{\hbar c /e B}$ is much larger than the $C$-$C$ distance times
the nanotube radius $R$, the low-energy theory is given by the linear
coupling of a two-component Dirac spinor to the corresponding vector
potential. We investigate in this regime the evolution of the
band structure of zig-zag nanotubes for values of $R/l > 1$, showing
that for radius $R \approx 20$ nm a clear pattern of Landau levels start
to develop for magnetic field strength $B \gtrsim 10$ T. The levels
tend to be four-fold degenerate, and we clarify the transition to
the typical two-fold degeneracy of graphene as the nanotube is
unrolled to form a curved strip. We show that the dynamics of the
Dirac fermions leads to states which are localized at the flanks of
the nanotube and that carry chiral currents in the longitudinal
direction. We discuss the possibility to observe the quantization
of the Hall conductivity in thick carbon nanotubes, which should
display steps at even multiples of $2 e^2/h$, with values doubled
with respect to those in the odd-integer quantization of graphene.

\end{abstract}

\maketitle

\section{Introduction}

Two-dimensional carbon compounds with $sp^2$ bonding have attracted
recently much attention, due to the experimental observation of a number
of novel electronic properties. It has been possible to measure the
transport properties of a single layer of graphite (so-called graphene),
providing evidence that the quasiparticles have a conical dispersion
around discrete Fermi points\cite{novo,zhang}.
Carbon nanotubes can be also considered as the result of
wrapping up the graphene sheet, leading to systems with
unconventional transport properties that reflect the strong electron
correlation\cite{egger,kane}.

The metallic carbon nanotubes and the graphene sheet have in common
that their low-energy electronic dispersion is governed by a massless
Dirac equation, around each of the two Fermi points of the undoped
systems\cite{mele1,nos,mele2}.
The appearance of an additional pseudo-spin quantum number intrinsic
to the Dirac spectrum has allowed us to understand, for instance, the
degeneracy of the molecular orbitals in the fullerenes\cite{prl}, the
quantization rule of the Hall conductivity in graphene\cite{graph1,graph2},
or the properties of the polarizability in carbon nanotubes\cite{lev}.

In this paper we investigate the effects of a transverse magnetic field
on the transport properties of the carbon nanotubes, making
use of the description of the electronic states in terms of Dirac fermion
fields. The low-energy graphene band structure can be obtained by taking a
continuum limit in which the momenta are much smaller than the inverse
of the $C$-$C$ distance $a$ \cite{mele1,nos,mele2}. In the case of carbon
nanotubes under transverse magnetic field, a sensible continuum limit
requires also that the square of the magnetic length $l^2 = \hbar c /e B$
is made much larger than $a$ times the nanotube radius, so that lattice
effects can be disregarded. In that limit, we will obtain a simple field
theory of Dirac spinors coupled to the magnetic field, allowing us to
investigate the dependence of different features of the band structure on
the topology of the space.

We will see that carbon nanotubes of sufficiently large radius
may have a quantum Hall regime, with a quantized Hall conductivity
$\sigma_{xy}$. In the case of graphene, it has been shown that
$\sigma_{xy}$ has plateaus at odd multiples of $2 e^2 /h $\cite{novo,zhang},
as a consequence of the peculiar Dirac spectrum\cite{graph1,graph2}. We will
find that the different topology of the carbon nanotubes leads instead to a
quantization in even multiples of the quantity $2 e^2 /h $, with steps in
$\sigma_{xy}$ which are doubled with respect to those in graphene.
We will also show how the transition to the odd-integer quantization of
graphene takes place, as the nanotube is unrolled into a curved strip.

\section{Continuum limit of carbon nanotubes in transverse magnetic
field}

\begin{figure}
\begin{center}
\mbox{\epsfxsize 5cm \epsfbox{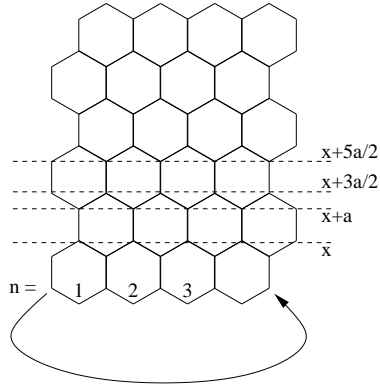}}
\end{center}
\caption{Schematic representation of the lattice of a zig-zag nanotube,
showing the four different levels of inequivalent atoms in a unit cell.}
\label{zigzag}
\end{figure}

In this section we show how to take the continuum limit of carbon
nanotubes, when the relevant electronic excitations have a wavelength
much larger than the $C$-$C$ distance $a$.
We illustrate this long-wavelength limit in the case of zig-zag
nanotubes, noting that the procedure works similarly for different
helicities. The zig-zag nanotubes have a unit cell with length $3a$,
containing four transverse arrays of carbon atoms at different longitudinal
positions $ x_j = 0, a, 3a/2$ and $5a/2$, as shown in Fig. \ref{zigzag}.
This introduces a flavor index $j = 1, \ldots 4$ labeling inequivalent
atoms in the unit cell.
It is convenient to introduce the Fourier transform of the electron
operator $\Psi (x_j + 3ma, n)$ with respect to the position
of the carbon atoms $n = 1,2, \ldots N$ in each transverse section
\begin{equation}
\Psi (x_j + 3ma , n) \sim \sum_p  e^{i 2\pi np/N} \; \Psi_j (m,p)
\end{equation}
where $m \in Z$ runs over the different cells.
The index $p$
labels then the different 1D subbands, $p = 0, \pm 1, \ldots
\pm (N-1)/2$ (or $N/2$) for the case of odd (even) $N$.

Within the tight-binding approach, the hamiltonian for the
zig-zag nanotube (without magnetic field) is given by
\begin{eqnarray}
 H_{tb}  &  =  &     - t
           \sum_{p,m} \Psi^+_1 (m,p) \Psi_2 (m,p)
                                       \nonumber   \\
      &    &  - t  \sum_{p,m}  z_p
       \Psi^+_2 (m,p) \Psi_3 (m,p)
                                \nonumber   \\
  &    &  - t  \sum_{p,m}
       \Psi^+_3 (m,p)  \Psi_4 (m,p)
                                 \nonumber      \\
  &   &   - t   \sum_{p,m}  z^{*}_p
     \Psi^+_4 (m,p) \Psi_1 (m+1,p) \;\;\;  + \;\;\;  {\rm h. c.}
\label{tb}
\end{eqnarray}
where $t$ is the hopping integral and $z_p \equiv
1 +  \exp (i 2\pi p/N  )$.
In the absence of magnetic field, the different subbands labeled
by $p$ are decoupled and, after introducing the longitudinal
momentum $k$, the hamiltonian corresponds to a $4 \times 4$ matrix
describing the unit cell:
\begin{equation}
\left.  {\cal H}_{p,p'} \right|_{B=0}   =
 -  \delta_{p, p'} \; t  \left(
\begin{array}{cccc}
  0 & 1 & 0 & e^{ -i 3 k a  } z_p \\
  1 & 0 & z_p &  0 \\
  0 & z^*_p & 0 & 1  \\
  e^{ i 3 k a  } z^*_p  & 0 & 1 & 0 \\
\end{array}
       \right)
\label{h44}
\end{equation}

The diagonalization of (\ref{h44}) leads in general
to massive subbands with parabolic dispersion, with a gap $2
\Delta_p = 2 t | 1 - 2 \cos (  \pi p / N ) |$. We note that,
whenever $N$ is a multiple of 3, the gap vanishes for $p= \pm N/3$
and we get two different couples of massless branches crossing in
each case at $k = 0$ but with opposite angular momentum around
the nanotube.
In general, the dispersive branches can be decoupled from
the high-energy branches that appear near the top of the spectrum. It
turns out that the low-energy dispersion corresponds to two reduced
2-component spinors described each by the $2 \times 2$ hamiltonian
\begin{equation}
\left.  {\cal H}_{p,p'} \right|_{B=0}   =
  \delta_{p, p'}  \left(
\begin{array}{cc}
v_F \hbar k     &    \Delta_p     \\
\Delta_p      &    - v_F \hbar k
\end{array}         \right)
\label{part1}
\end{equation}
where the Fermi velocity is $v_F = 3ta/2\hbar $. We see that the two
components of each spinor, that we will denote by  $\Psi_R$ and $\Psi_L$,
correspond respectively to right- and left-moving modes, which are mixed by
the massive (off-diagonal) term of the hamiltonian.

The magnetic field is introduced with the usual prescription of
correcting the transfer integral $t$ by appropriate phase factors
\begin{equation}
e^{i \phi }  =
\exp \left(i \frac{e}{\hbar c}
    \int_{\bf r}^{\bf r'} {\bf A} \cdot {\bf d l} \right)
\label{exp}
\end{equation}
depending on the vector potential ${\bf A}$ between
nearest-neighbor lattice sites at ${\bf r}$ and ${\bf r'}$. For a
magnetic field perpendicular to the carbon nanotube, the component
normal to the surface
has a periodic dependence on the angular variable $\theta $
around the tube. Taking the longitudinal distance $x$ and the
angle $\theta $ as coordinates on the nanotube surface, a convenient
choice of the gauge is given by
\begin{equation}
{\bf A} = (RB \sin (\theta ), 0)
\end{equation}
$R$ being the nanotube radius. The phase $\phi $
gets then a modulation around the nanotube\cite{dress}
\begin{equation}
   \phi  \propto
   a (e/\hbar c) BR \sin (2\pi n/N)  \;\;\;\;\;\;    n = 1, \ldots N
\label{phase}
\end{equation}

The feasibility
of the continuum limit requires that $(e/\hbar c) BRa \ll 1$. When
this condition is satisfied, we can deal with the linear approximation
to the exponential (\ref{exp}). We observe then that the effect of
the magnetic field is to mix a given subband $p$ with its nearest
neighbors $p \pm 1$. This introduces another contribution to add to the
hamiltonian (\ref{h44}), given by
\begin{eqnarray}
&& \Delta {\cal H}_{p,p'}    =
  \delta_{p', p + 1}  \; t \frac{eBRa}{2\hbar c}  \left(
\begin{array}{cccc}
  0 & 1 & 0 & -e^{ -i 3 k a  } z_{p+1}/2 \\
  -1 & 0 & z_p/2 &  0 \\
  0 & -z^*_{p+1}/2 & 0 & 1  \\
  e^{ i 3 k a  } z^*_p/2  & 0 & -1 & 0 \\
\end{array}
 \right)  \nonumber \\
&& + \delta_{p', p - 1}  \; t \frac{eBRa}{2\hbar c}  \left(
\begin{array}{cccc}
  0 & -1 & 0 & e^{ -i 3 k a  } z_{p-1}/2 \\
  1 & 0 & -z_p/2 &  0 \\
  0 & z^*_{p-1}/2 & 0 & -1  \\
  -e^{ i 3 k a  } z^*_p/2  & 0 & 1 & 0 \\
\end{array}
 \right)
\end{eqnarray}
By projecting again onto the two-dimensional low-energy space,
$\Delta {\cal H}_{p,p'} $ becomes
\begin{equation}
\Delta {\cal H}_{p,p'}    =
  \delta_{p', p \pm 1}  \left(
\begin{array}{cc}
  \pm i  v_F (e/c) BR/2     &    0    \\
   0    &    \mp  i  v_F (e/c) BR/2
\end{array}         \right)
\label{part2}
\end{equation}

The hamiltonian can be more easily expressed when acting on the
two-component Dirac spinor
\begin{equation}
\Psi (k, \theta ) \sim \sum_p e^{i \theta p} \Psi (k,p)
\end{equation}
depending on the angular variable
$\theta $ around the tubule. We recall that there are in fact two
different spinors describing states with opposite angular momentum
around the nanotube. For a nanotube without gap, for instance, the
hamiltonian is in either case
\begin{equation}
{\cal H} =  \left(
\begin{array}{cc}
 v_F \hbar k  +  v_F \frac{eBR}{c} \sin (\theta )  &  -i(\hbar v_F /a ) \partial_{\theta }  \\
 -i(\hbar v_F /a ) \partial_{\theta }  &  - v_F \hbar k  -  v_F \frac{eBR}{c} \sin (\theta )
\end{array}
   \right)
\label{dirac}
\end{equation}
where the periodic modulation matches with the orientation of the
magnetic field normal to the nanotube surface at $\theta = 0$.
Expression (\ref{dirac}) corresponds actually to the Dirac
hamiltonian with the usual prescription for the coupling to the
vector potential,
$\hbar k \rightarrow \hbar k + (eBR/c) \sin (\theta )$. We will
see, however, that this simple gauge coupling does not hold
in all cases, when discussing the curved graphene strips in the
next section.

\section{Landau level quantization}

\begin{figure}
\begin{center}
\mbox{\raisebox{2.0cm}{$\varepsilon $} \epsfxsize 5cm
\epsfbox{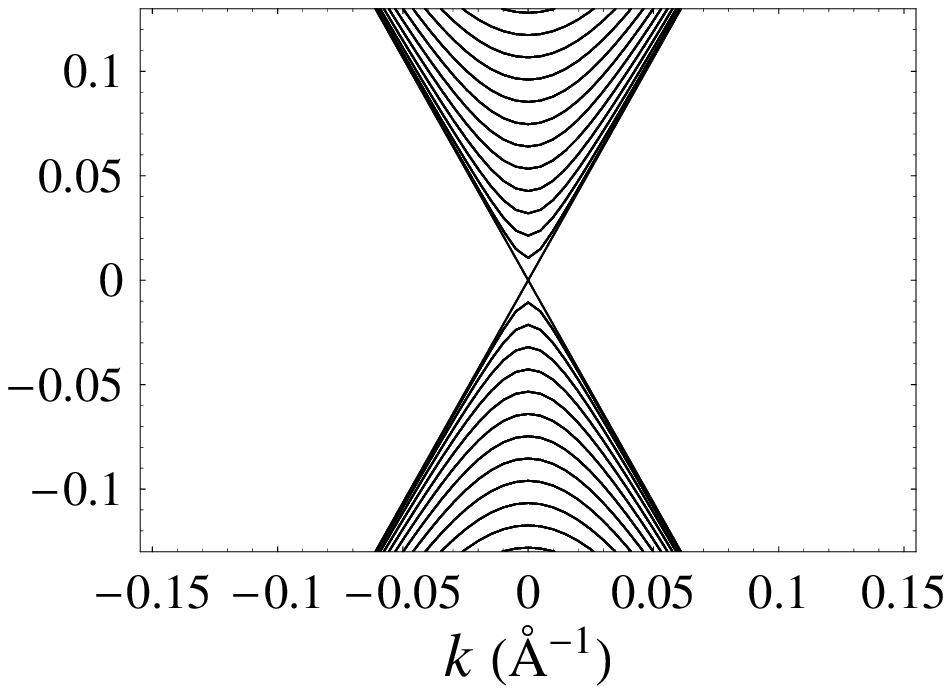} \raisebox{2.0cm}{$\varepsilon $} \epsfxsize
5cm
\epsfbox{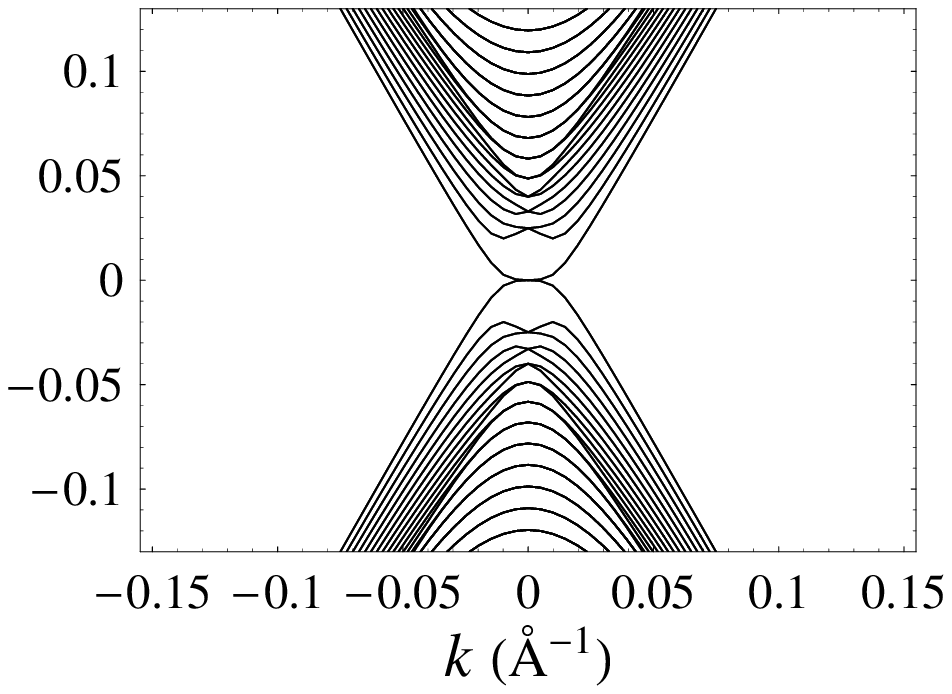}  }\\
 \hspace{0.65cm}  (a) \hspace{4.75cm} (b)   \\   \mbox{}  \\
\mbox{\raisebox{2.0cm}{$\varepsilon $} \epsfxsize 5cm
\epsfbox{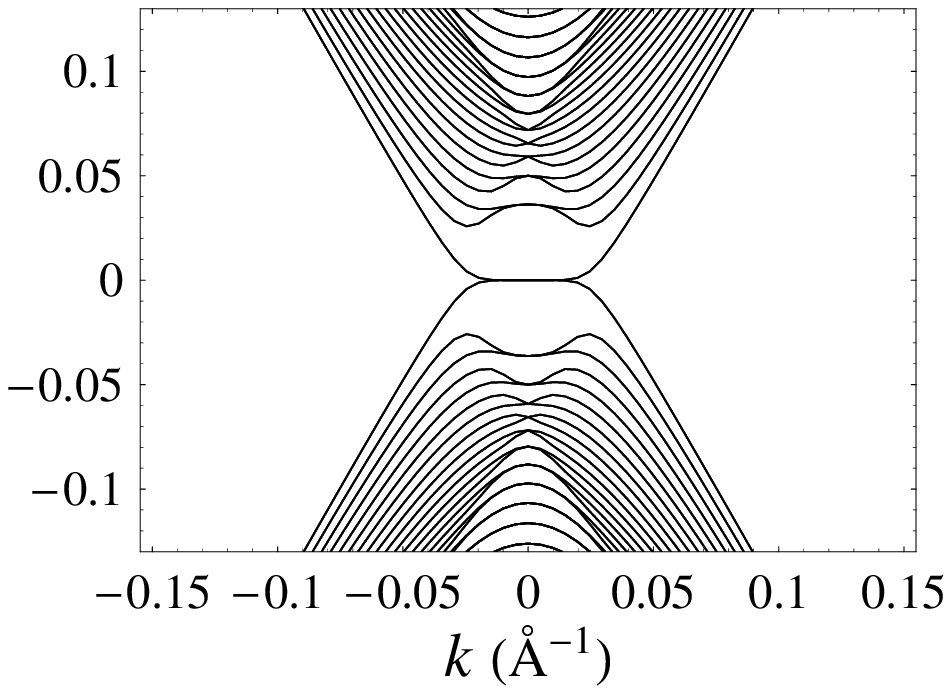} \raisebox{2.0cm}{$\varepsilon $} \epsfxsize
5cm
\epsfbox{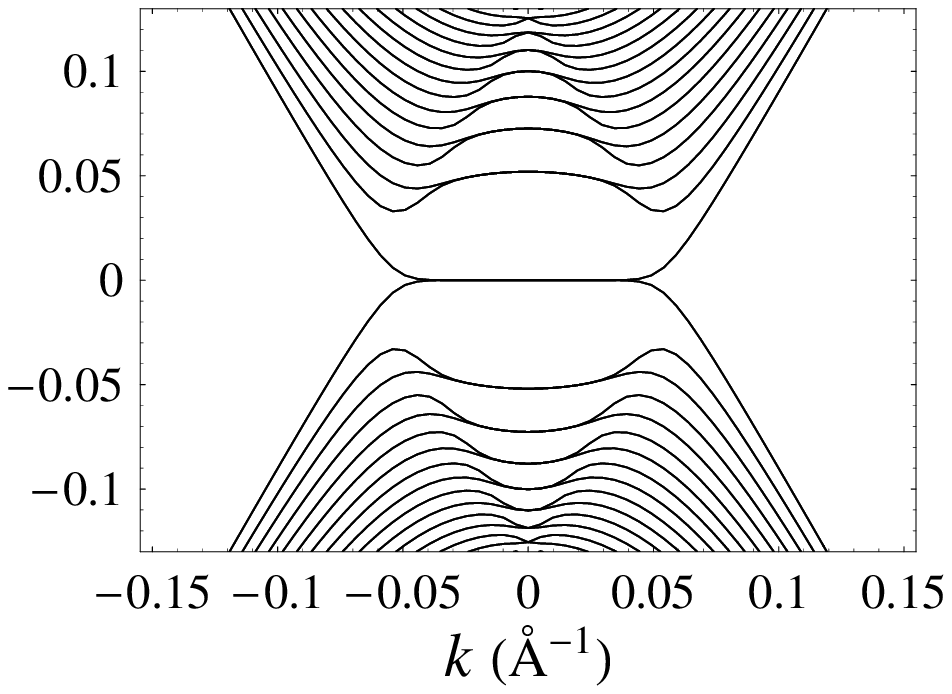} }\\
 \hspace{0.65cm}  (c) \hspace{4.75cm} (d)
\end{center}
\caption{Sequence of band structures of a zig-zag (510,0) nanotube
with radius $R \approx 20$ nm in transverse magnetic field, for
$B=0$ T (a); $B=5$ T (b); $B=10$ T (c); $B=20$ T (d). $B=20$ T
corresponds to $aR/l^{2} \approx 0.1$ and $R/l \approx 3.5$.
Energy is in units of $t$ and momentum is in units of \AA$^{-1}$.}
\label{met}
\end{figure}

\begin{figure}
\begin{center}
\mbox{\raisebox{2.0cm}{$\varepsilon $}
\epsfxsize 5cm \epsfbox{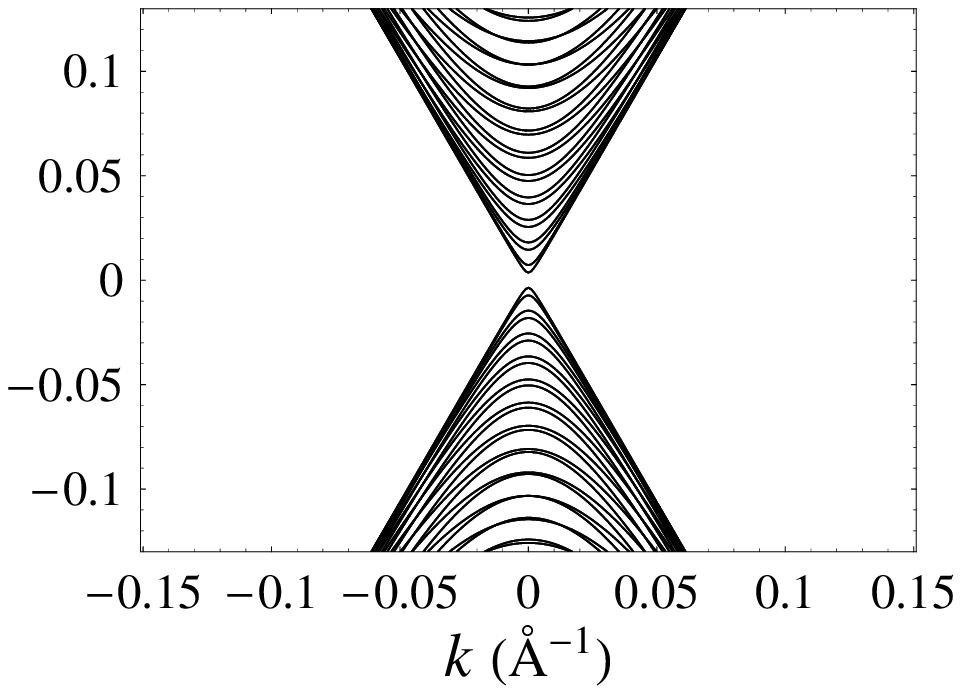}
\raisebox{2.0cm}{$\varepsilon $} \epsfxsize 5cm
\epsfbox{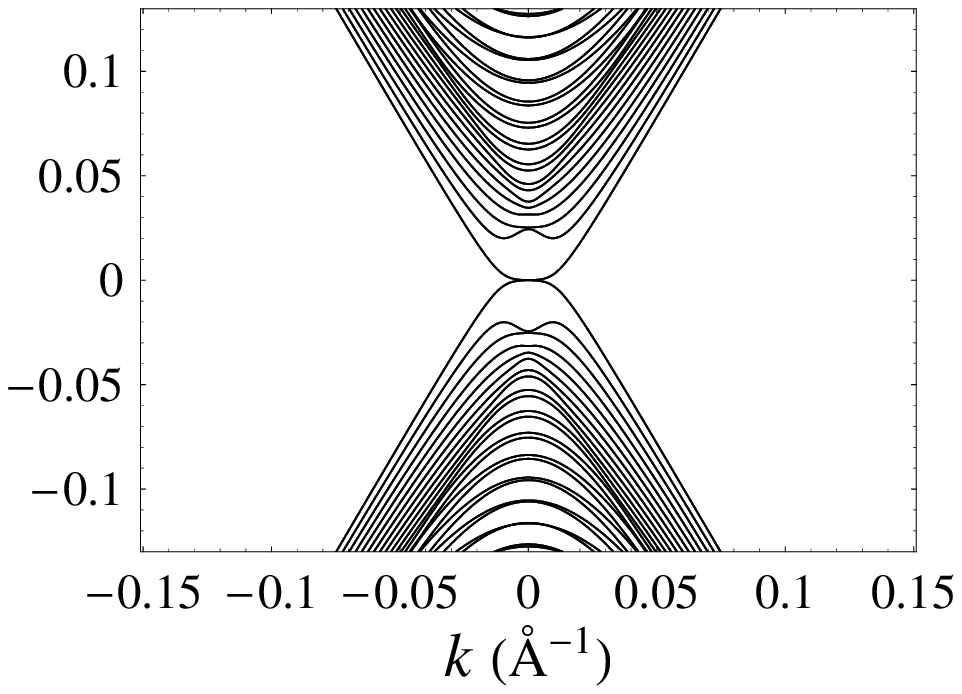}  }\\
 \hspace{0.65cm}  (a) \hspace{4.75cm} (b)   \\   \mbox{}  \\
\mbox{\raisebox{2.0cm}{$\varepsilon $}
\epsfxsize 5cm \epsfbox{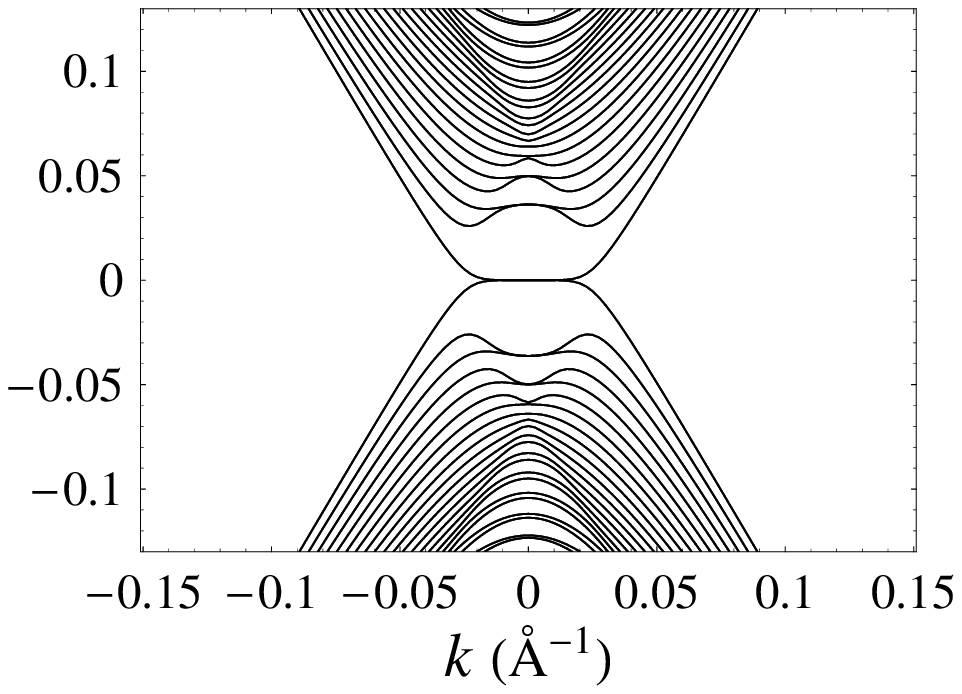}
\raisebox{2.0cm}{$\varepsilon $} \epsfxsize 5cm
\epsfbox{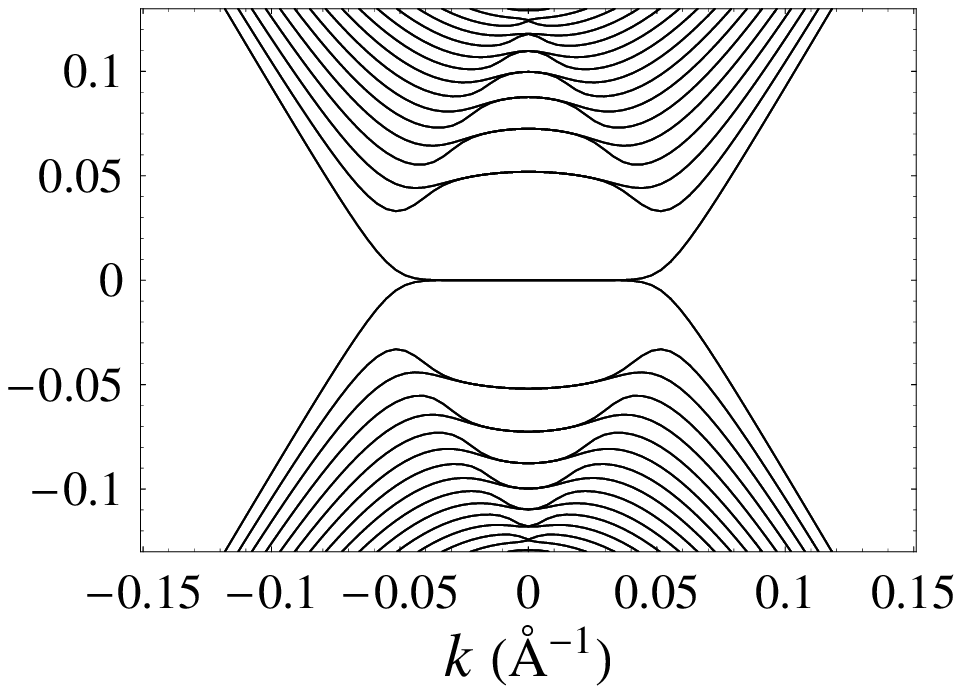} }\\
 \hspace{0.65cm}  (c) \hspace{4.75cm} (d)
\end{center}
\caption{Sequence of band structures of a zig-zag (500,0) nanotube
with radius $R \approx 20$ nm in transverse magnetic field, for
$B=0$ T (a); $B=5$ T (b); $B=10$ T (c); $B=20$ T (d).
Energy is in units of $t$ and
momentum is in units of \AA$^{-1}$.}
\label{sem}
\end{figure}

We have diagonalized numerically the hamiltonian ${\cal H}$ made of
the sum of (\ref{part1}) and (\ref{part2}),
for different carbon nanotubes with radius $R \approx 20$ nm and
magnetic field $B$ varying between 0 T and 20 T.
We have checked that, for $a R/l^2 \ll 1$, the eigenstates of
${\cal H}$ provide a good approximation to the low-energy
band structure obtained from the full tight-binding hamiltonian
incorporating the phase factors (\ref{exp}) for the different bonds.
We present our results showing the evolution of the band structure
computed from the hamiltonian ${\cal H}$
for two different zig-zag nanotubes, corresponding to (510,0) and
(500,0) geometries in the usual notation. We observe that, while the
latter has a small gap in the absence of magnetic field, the
evolution represented in Figs. \ref{met} and \ref{sem} ends up
in quite similar shapes for the band structure at strong magnetic
field ($\approx 20$ T). The closure of the gap at a magnetic field
$B < 5$ T is consistent with the results for semiconducting carbon
nanotubes in Ref. \onlinecite{lu}. From there we can infer that
the magnetic field needed to close the gap of a nanotube with radius
$R = 10$ nm must be of the order of $\sim 10$ T. This field strength
would be reduced by a factor of 4 after doubling the nanotube radius,
keeping the same ratio of $R/l$.
For the larger radius $R \approx 20$ nm in our analysis, we find that
the band structures at $B = 5$ T in Figs. \ref{met}(b) and \ref{sem}(b)
only differ in the position of some unpaired subbands
in the latter figure. At $B = 10$ T a discrepancy between
Figs. \ref{met}(c) and \ref{sem}(c) is only found in the high-energy
part of the plots, and at $B = 20$ T the band structures for the
metallic and the semiconducting nanotube represented in
Figs. \ref{met}(d) and \ref{sem}(d) are practically similar.
This illustrates a more general result,
which is that the form of the low-energy Landau subbands does not
depend on the particular geometry of nanotubes with similar radius
at such strong magnetic fields.

We have to point out however that the evolution of the band structure
of the thick nanotubes considered here is quite different from that of
carbon nanotubes with typical radius ($\sim 1$ nm) in strong magnetic
fields. The latter have been investigated in Ref. \onlinecite{add},
where typical oscillations have been reported in the low-energy levels
of carbon nanotubes with $R \sim 1$ nm as the magnetic field is
increased to ratios of $R/l = 3$. The reason why the low-energy levels
do not stabilize at increasing magnetic field can be traced back to
the fact that, for such thin carbon nanotubes, there is no regime where
the continuum limit with $a R/l^2 \ll 1$ can be realized. In these cases,
by the time that we have $R \gtrsim l$, the magnetic length cannot be
much larger than the $C$-$C$ distance, so that a quantum Hall regime
cannot exist in thin carbon nanotubes of typical radius. This can be
also appreciated in the results of Ref. \onlinecite{cuni}, where the
density of states of several carbon nanotubes is represented at very
large magnetic fields, with a marked difference between the cases of
thin and thick nanotubes. It has been shown for instance that the
density of states for nanotube radius $R \approx 14$ nm already resembles
that of the parent graphene system, with clear signatures of Landau
subbands in the low-energy part of the spectrum.

As represented in Figs. \ref{met} and \ref{sem},
our thick nanotubes develop in general two valleys at
zero energy (that appear superposed in the figures) expanding
around the two Fermi points of the parent graphene system at $B =0$.
We see that flat Landau levels start developing already at $B=10$ T
(Fig. \ref{met}(c)).
The existence of a zero-energy
level at $k = 0$ has been shown to be a robust property of
carbon nanotubes in a transverse magnetic field\cite{ando,novikov}.
We have checked that, for large magnetic field strength
(as in Fig. \ref{met}(d)), the energy levels at $k = 0$
follow the quantization rule $\varepsilon_n \propto \sqrt{n}$ \cite{novikov},
which is peculiar of graphene\cite{mac,dress}.   The
point that we want to stress here is our observation that the
levels at $k = 0$ are four-fold degenerate, including the
zero-energy level, for any kind of nanotube geometry.
This is in contrast with the case of planar graphene, where the
zero-energy level is doubly degenerate. The reduction in the
number of zero modes comes from the fact that the boundary conditions in
the plane impose in general the hybridization of states with opposite
transverse momentum, while they are otherwise independent in the tubular
geometry.
As we will
see, this bears a direct relation to the quantization of the Hall
conductivity in even multiples of $2e^2 /h$ in the carbon nanotubes.


\begin{figure}
\begin{center}
\mbox{\epsfxsize 7.5cm \epsfbox{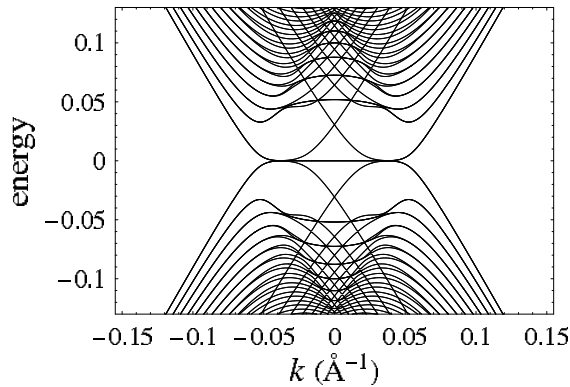}}
\end{center}
\caption{Band structure corresponding to a zig-zag nanotube with
a cut in the longitudinal direction, for $R \approx 20$ nm and
$B = 20$ T (in the same units as in Fig. \ref{met}).}
\label{two}
\end{figure}

A relevant question is how the four-fold degeneracy of the above band
structure can evolve into a picture consistent with the odd-integer
quantization of the Hall conductivity in the planar graphene samples.
This can be clarified by studying the change in the band structure after
cutting the nanotube along the longitudinal direction.
If we cut along a maximum in the normal component of
the magnetic field, this leads to a strip with a dependence of the
tight-binding phase as in (\ref{phase})
\begin{equation}
\phi \propto a (e/\hbar c) BR \sin (2\pi n/N)
\end{equation}
We may obtain the band structure from the spectrum
of an imaginary nanotube made by matching two copies of the strip,
after projection onto the set of odd eigenmodes of the nanotube (which
correspond to the stationary waves in the transverse direction of the
strip\cite{brey}).
The phase $\phi $ in the imaginary nanotube has to keep the mirror symmetry
between the two copies. This leads to a function which does not have
continuous derivative at the matching point of the two strips, with an infinite
expansion in the modes around the nanotube. The corresponding interaction
term is
\begin{equation}
\Delta {\cal H}_{p,p'}    =
 \sum_m \delta_{p', p \pm 2m + 1} f(p-p')
\left(
\begin{array}{cc}
   v_F (e/c) BR/2    &    0    \\
   0    &    -  v_F (e/c) BR/2
\end{array}         \right)
\end{equation}
with $f(p) = - 8/\pi (p^2 - 4)$.
The band structure obtained after projection of the spectrum of
${\cal H}$ onto the odd eigenmodes of the nanotube is represented in
Fig. \ref{two}. We observe that states that were at zero energy in the
plot of Fig. \ref{met}(d) form two subbands dispersing towards higher energies,
and two other dispersing downwards. In these conditions, only one valley
is left at zero energy, evidencing the transition to the odd-integer
quantization of the Hall conductivity of graphene.

A more complete picture of this transition can be obtained by studying
an intermediate situation, where the nanotube has been cut longitudinally
as before and opened by an angle of $\pi $ in the transverse magnetic field.
This case is simple to analyze,
since it corresponds to a strip with a modulation of the tight-binding
phase given by
\begin{equation}
\phi \propto a (e/\hbar c) BR \cos (\pi n/N)
\label{cos}
\end{equation}
We may
form a nanotube by matching two copies of the strip, obtaining
again the band structure of the latter by projection of the spectrum
of the nanotube onto eigenmodes with odd angular dependence.
Now the doubling of (\ref{cos}) realizing the mirror symmetry between
the two copies gives back the same function $\phi $,
and the interaction term becomes
\begin{equation}
\Delta {\cal H}_{p,p'}    =
  \delta_{p', p \pm 1}
\left(
\begin{array}{cc}
  v_F (e/c) BR/2     &    0    \\
   0    &    -   v_F (e/c) BR/2
\end{array}         \right)
\end{equation}
The band structure obtained from the diagonalization of the full
hamiltonian ${\cal H}$ has a shape similar to that shown in
Fig. \ref{met}. The important difference is that now the projection onto
the odd eigenmodes of the nanotube reduces by a factor of 2 the number
of states, leaving all the subbands non-degenerate. We find therefore
that, in the curved strip which is formed by opening the nanotube by an
angle of $\pi $, the zero-energy level is already two-fold degenerate,
while the rest of subbands coalesce in pairs at low momentum into
two-fold degenerated levels. This is actually the kind of
degeneracy that leads to the odd-integer quantization of the Hall
conductivity in the planar graphene samples\cite{graph1,graph2}.

>From the above results we may also infer that the band structure of the
carbon nanotube can be obtained from the superposition of the subbands
of two nanotube halves in the transverse magnetic field. We
can think of these two curved strips as the upper and the lower half
of the carbon nanotube, leading to a picture in which gluing or not
the two strips should be immaterial for the purpose of reproducing
the nanotube band structure. It can be shown that this is actually the
case, by looking at the result of cutting the nanotube at one of the
flanks, where the normal component of the magnetic field vanishes.
Unrolling then the nanotube leads to a strip with a modulation
of the tight-binding phase
\begin{equation}
\phi \propto a (e/\hbar c) BR \cos (2\pi n/N)
\end{equation}
Its band structure can be deduced again from that of an imaginary nanotube
made by matching two copies of the strip, with the same phase $\phi$, after
projection onto the set of odd eigenmodes of the nanotube. This nanotube
model has an interaction term
\begin{equation}
\Delta {\cal H}_{p,p'}    =
  \delta_{p', p \pm 2}
\left(
\begin{array}{cc}
  v_F (e/c) BR/2     &    0    \\
   0    &    -  v_F (e/c) BR/2
\end{array}         \right)
\end{equation}
The full hamiltonian ${\cal H}$ has now an even dependence on the quantum
number $p$, and it can be checked that the result of projecting the
spectrum onto the eigenmodes with odd angular dependence produces
in fact a band structure which is completely similar to that in
Fig. \ref{met}. This provides the final evidence that the doubling of
the spectrum of the carbon nanotube can be understood precisely
from the doubling of the geometry in two equivalent upper and lower
nanotube halves, which contribute equally to the spectrum in the
transverse magnetic field.

\section{Current quantization}

We clarify next
the way in which the dispersive branches of the
band structure are connected to the quantization of the Hall
conductivity in the carbon nanotubes. The outer dispersive branches
correspond to states which are localized at the flanks of the
nanotube\cite{ando}.
This suggests the possibility that, despite having no boundary,
the carbon nanotubes may support edge states in similar fashion as
in systems with planar geometry. It can be shown actually that the
current carried in the longitudinal direction by the states in the
outer dispersive branches is quantized. In this context, the definition
of the current $j$ must be consistent with the dynamics governed by
the hamiltonian (\ref{dirac}). In real space the equation of motion
for the Dirac spinor $\Psi (x, \theta )$ reads
\begin{equation}
i \hbar \partial_t
\left(
\begin{array}{c}
\Psi_R       \\
\Psi_L
\end{array}
   \right)  =
\left(
\begin{array}{cc}
 -i v_F \hbar \partial_x  +  v_F \frac{eBR}{c} \sin (\theta )
                           &  -i(\hbar v_F /a ) \partial_{\theta }  \\
 -i(\hbar v_F /a ) \partial_{\theta }
             &   i v_F \hbar \partial_x  -  v_F \frac{eBR}{c} \sin (\theta )
\end{array}
   \right)
\left(
\begin{array}{c}
\Psi_R       \\
\Psi_L
\end{array}
   \right)
\label{eqm}
\end{equation}
>From (\ref{eqm}) we derive the continuity equation
\begin{equation}
\partial_t (\Psi_R^{+} \Psi_R + \Psi_L^{+} \Psi_L )
   = - v_F \partial_x (\Psi_R^{+} \Psi_R - \Psi_L^{+} \Psi_L )
\end{equation}
which dictates the expression of the current
\begin{equation}
j = v_F (\Psi_R^{+} \Psi_R - \Psi_L^{+} \Psi_L )
\label{cur}
\end{equation}

\begin{figure}
\begin{center}
\mbox{\epsfxsize 7.5cm \epsfbox{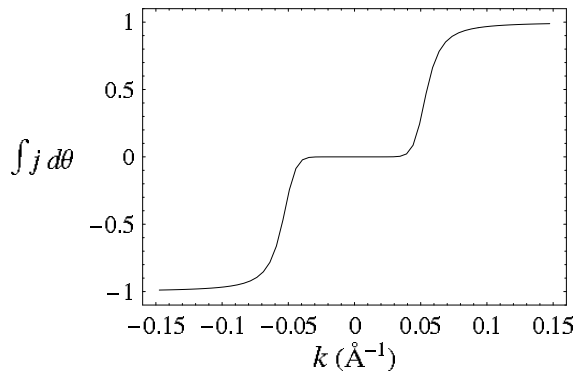}}
\end{center}
\caption{Plot of the integral of the current $j$ over the angular
variable $\theta $ (in units of $v_F$)
for states in the lowest Landau subband of Fig. \ref{met}(d).}
\label{three}
\end{figure}

The result of computing the integral over $\theta $ of the current $j$
for states in the lowest
energy subbands is represented in Fig. \ref{three}. It turns out that,
in general, the states corresponding to the flat part of the
Landau level do not carry any current in the longitudinal direction,
while the states in the dispersive branches saturate quickly the
maximum value $v_F$ as the dispersion approaches a constant slope.

\begin{figure}
\begin{center}
\mbox{\raisebox{1.8cm}{$|\Psi_R|^2$}
\epsfxsize 4.2cm \epsfbox{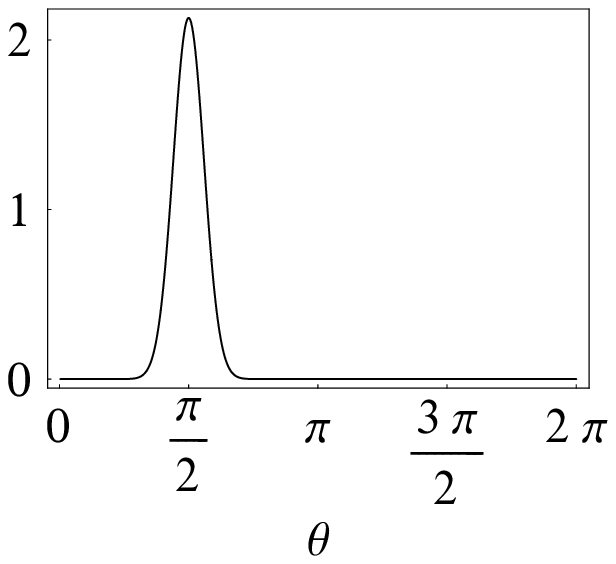}
\raisebox{1.8cm}{$|\Psi_L|^2$} \epsfxsize 4.2cm
\epsfbox{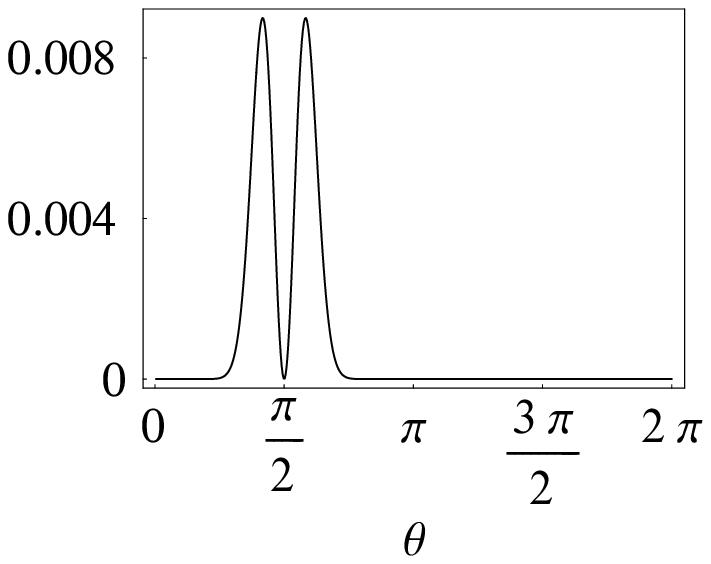}  }\\ \mbox{} \\
\mbox{\raisebox{1.8cm}{$|\Psi_R|^2$}
\epsfxsize 4.2cm \epsfbox{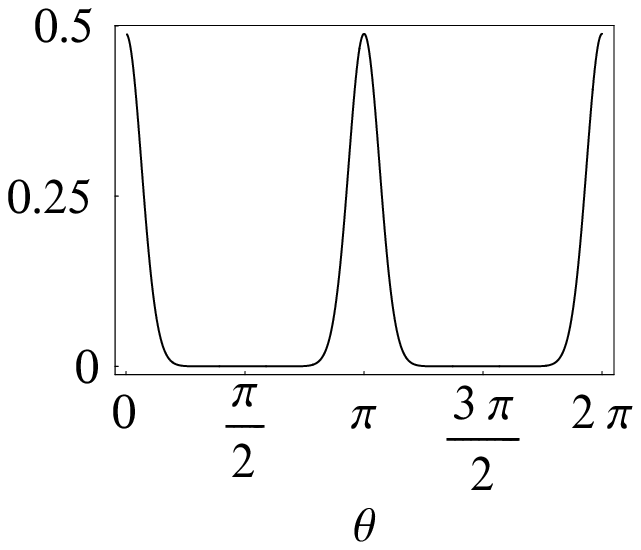}
\raisebox{1.8cm}{$|\Psi_L|^2$} \epsfxsize 4.2cm
\epsfbox{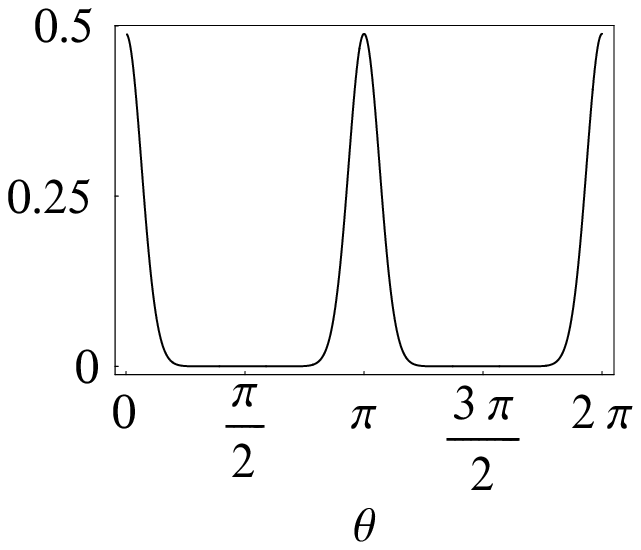} } \\ \mbox{} \\
\mbox{\raisebox{1.8cm}{$|\Psi_R|^2$}
\epsfxsize 4.2cm \epsfbox{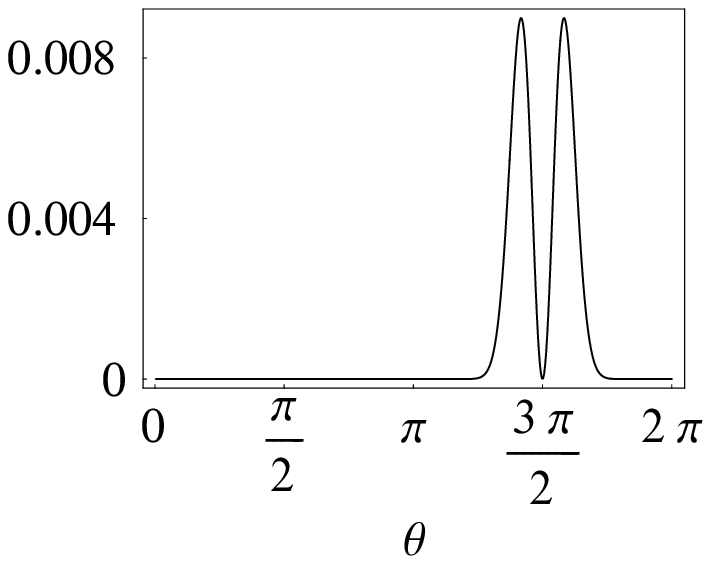}
\raisebox{1.8cm}{$|\Psi_L|^2$} \epsfxsize 4.2cm
\epsfbox{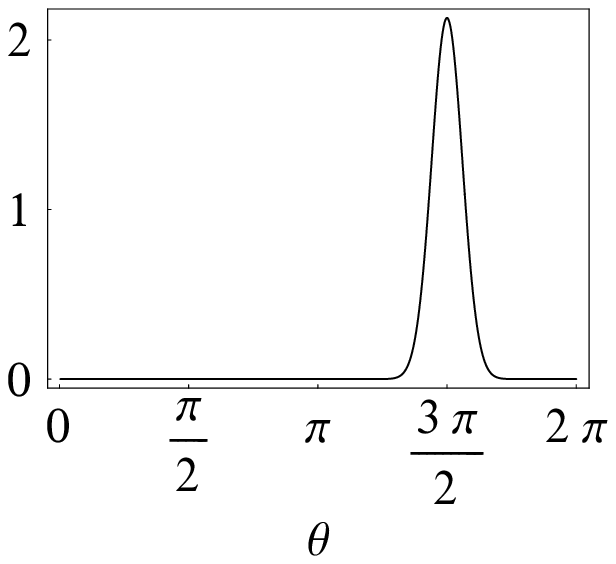} }
\end{center}
\caption{Angular distribution of the eigenfunctions in the lowest
Landau subband of Fig. \ref{met}(d). The panels correspond, from top
to bottom, to longitudinal momenta $k = 0.15, 0, -0.15$ (in
units of \AA$^{-1}$).}
\label{eigen}
\end{figure}

Regarding the spatial distribution, there is also a clear correspondence
between the localization of the states in the angular variable $\theta $
and the value of the current. This can be appreciated from inspection
of the eigenstates of the hamiltonian (\ref{dirac}). We have represented
in Fig. \ref{eigen} the angular distribution of states from the lowest
Landau subband  for $B = 20$ T. Each
eigenfunction is in general localized around a certain value of the
angular variable $\theta $. We observe, for instance, that the states
at $k = 0$ have gaussian wave functions localized at $\theta = 0$
or $\theta = \pi $, with the contribution to the current from the
left component compensating exactly that from the right component.
For positive (negative) longitudinal momentum, the states in the flat
zero-energy level
are localized at angles between 0 and $\pi /2 $ ($3\pi /2
$), or between $\pi $ and $\pi /2 $ ($3\pi /2 $), depending on the
subband chosen. For the states in the dispersive branches, the
eigenfunctions are centered around $\pi /2 $ (for a right branch) or
$3\pi /2 $ (for a left branch). Here the role of the magnetic field
is to separate left-moving and right-moving currents at opposite sides of
the tube. We actually observe that there is a large mismatch between
$|\Psi_L |$ and $|\Psi_R |$ for states in the dispersive branches, which
turn out to carry nonvanishing chiral currents flowing at the
flanks of the nanotube.

The localization of the current in the states of the dispersive
branches opens the possibility to observe the quantization
of the Hall conductivity in thick carbon nanotubes.
We first remark that, in general, the value of the current $j$ (more
precisely the integral over $\theta $) is
proportional to the slope of the dispersion $\varepsilon (k)$
in the given subband. This can be seen by taking the derivative
with respect to $k$ of the expectation value of (\ref{dirac}) in the
energy eigenstates, after use of the Hellmann-Feynman theorem:
\begin{equation}
\frac{1}{\hbar} \frac{\partial }{\partial k} \langle {\cal H} \rangle
 = \frac{1}{\hbar} \langle \frac{\partial }{\partial k} {\cal H} \rangle
 = v_F \int d \theta \; (\Psi_R^{+} \Psi_R - \Psi_L^{+} \Psi_L )
\end{equation}
Thus we obtain
\begin{equation}
 \frac{1}{\hbar}  \frac{\partial \varepsilon (k)}{\partial k }
      =     \int d \theta \; j
\label{hf}
\end{equation}

Let us consider first the case in which the Fermi level is above
the zero-energy level but without crossing the next Landau subband.
If we denote by $\varepsilon_0 (k)$ the dispersion of the lowest
Landau subband, we have from (\ref{hf}) that the total longitudinal
current $I$ is given then by the sum of
$(e/\hbar) \partial \varepsilon_0 (k) / \partial k $ over all the filled
modes in the energy range between the respective chemical potentials
$\varepsilon_L $ and $\varepsilon_R $ at the two nanotube flanks \cite{halp}
\begin{equation}
I = \frac{e}{\hbar} \int_{{\rm filled}\; {\rm states}} \frac{dk}{2\pi }
    \frac{\partial \varepsilon_0 (k) }{\partial k}
\end{equation}
Provided that the Hall probes used
to measure the difference in the chemical potential are noninvasive,
as proposed below, the Hall voltage $V_H$ measured by the probes has to
be given by $(\varepsilon_R - \varepsilon_L )/e$ \cite{ando2}.
Therefore the Hall conductivity, defined by
$\sigma_{xy} = I/V_H$, must have a first
plateau as a function of the filling level, with a
quantized value given by the spin degeneracy and the doubling of the
subbands shown in Fig. \ref{met}(d):
\begin{equation}
\sigma_{xy} = 4 \frac{e^2}{h}
\end{equation}

As the filling level is increased, the situation changes when the Fermi
level starts crossing the bumps with parabolic dispersion shown in Fig.
\ref{met}(d). The current given by Eq. (\ref{cur}) is again proportional
to the slope of the energy dispersion. Thus, when the Fermi level intersects
a parabolic branch, there are two opposite contributions to the
longitudinal current $I$ from the respective Fermi points. As the filling
level is further increased, however, the Fermi points move far apart, and
the contributions to the current $I$ depend on the profile of the Hall
potential across the section of the nanotube.

\begin{figure}
\begin{center}
\mbox{\epsfxsize 7.5cm \epsfbox{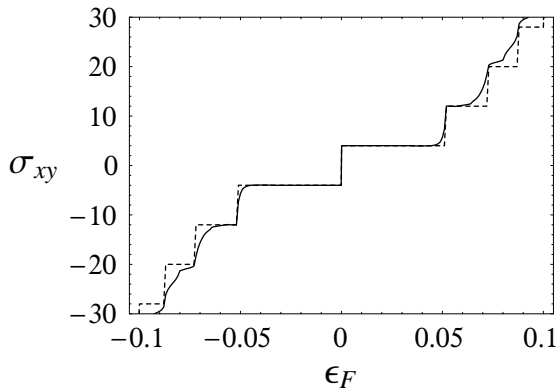}}
\end{center}
\caption{Plot of the Hall conductivity $\sigma_{xy}$ (in units of $e^2 /h$)
as a function of the position of the Fermi level $\varepsilon_F$ in the band
structure of Fig. \ref{met}(d). The full line corresponds to a linear Hall
voltage drop and the dashed line to a sharp voltage drop in the bulk of the
nanotube.}
\label{four}
\end{figure}

When the Fermi level crosses several of the dispersive branches shown
in Fig. \ref{met}(d), there is the question of how the total current $I$
is distributed among the different subbands. This problem has been
addressed in the context of the quantum Hall effect in quantum wires,
in experimental studies measuring the spatial profile of the Hall
potential\cite{vk}. It has been shown that such a spatial dependence
is dictated by the compressibility of the electron liquid. In the case
of the carbon nanotubes, the key point is the very limited screening
afforded by such low-dimensional systems. It seems therefore reasonable
to assume that the Hall potential drop is going to be approximately linear
in the bulk of the nanotube\cite{ando2}. In these conditions, we have
computed the Hall conductivity as a function of the filling level.
For the inner Landau subbands in Fig. \ref{met}(d), the difference in
the chemical potential between right and left branches
$\varepsilon_R - \varepsilon_L $ is in general below the
potential measured by the Hall contacts. The contribution of each
inner dispersive branch to the Hall conductivity turns out to be then
smaller than the quantized value from the outermost edge states.
Consequently, an approximate quantization of $\sigma_{xy}$ is observed
above the first plateau, as shown in Fig. \ref{four},
with steps according to the degeneracy of the subbands:
\begin{equation}
\sigma_{xy} \approx (2+4n)  2 \frac{e^2}{h}
\label{sigma}
\end{equation}

The appearance of sharper steps in the Hall conductivity is
achieved in general in samples with larger transverse size,
and it can be also favored by the use of wider Hall contacts that induce a
smoother drop of the Hall voltage at the edges\cite{ando2}.
In practice, a real measure of the Hall conductivity may be then between
the two curves displayed in Fig. \ref{four}.

\section{Conclusion}

We have shown that, for thick carbon nanotubes in a
transverse magnetic field, the transport properties are governed by
the states localized at the flanks of the nanotube, which carry
quantized currents in the longitudinal direction.

For nanotubes with a radius $R \approx 20 \; {\rm nm}$,
in a magnetic field of $\approx 20 \; {\rm T}$, the band structure
already shows a clear pattern of Landau levels. This opens the possibility
of observing the quantization of the Hall conductivity in multi-walled
nanotubes, where typically only the outermost shell is contacted by
electrodes in transport experiments. In a suitable experimental setup,
a metallic gate should be prepared as part of the substrate, to be used as a
voltage probe contacting the bottom of the nanotube. The magnetic field
should be oriented parallel to the substrate and perpendicular to the
carbon nanotube. Then, by establishing a fixed current along the nanotube,
changes in the Hall voltage at the top of the nanotube have to be observed
upon variation of the magnetic field strength or the chemical potential
in the nanotube. Longitudinal currents of the order of $\sim 1 \; \mu {\rm A}$
are quite affordable and, according to (\ref{sigma}), they have to give rise
to Hall voltages of the order of $\sim 1 \; {\rm mV}$. A scanning tunneling
microscope (STM) tip may be
used to contact the top part of the nanotube and to measure the voltage
with respect to that of the metallic gate at the bottom. As the STM
device may easily appreciate differences of the order of
$\sim 10^{-4} \; {\rm eV}$, it has to be possible to observe at least the
first steps in the Hall voltage implied by the quantization rule
(\ref{sigma}).

The observation of the plateaus in the Hall voltage should be fairly
insensitive to the presence of moderate disorder in the nanotube samples,
as long as the effect rests on the existence of chiral currents at
opposite flanks of a nanotube. The overlap between states with
currents flowing in opposite direction is exponentially small,
so that the chiral currents cannot
suffer significant backscattering from impurities or lattice defects.
It is only at the electrodes, where the chiral currents meet,
that backscattering may appear. As usually done in the context of
the Hall effect in mesoscopic wires, this may be accounted for by means
of a suitable transmission coefficient, that would reflect as an additional
factor in the relation between the longitudinal current and the Hall
voltage\cite{butt}.

Finally, we remark that the absence of significant backscattering
interactions leads to good perspectives to measure the properties of a
robust chiral liquid at the flanks of the nanotube, which could be
accomplished in particular by means of scanning tunneling spectroscopy.

The financial support of the Ministerio
de Educaci\'on y Ciencia (Spain) through grants
FIS2005-05478-C02-01/02 and INFN05-14 is gratefully acknowledged.
F. G. acknowledges funding from the European
Union Contract 12881 (NEST).
S.B. and P.O. acknowledge the support of the grant 2006 PRIN "Sistemi
Quantistici Macroscopici-Aspetti Fondamentali ed Applicazioni di
strutture Josephson Non Convenzionali".
E. P. was also supported by INFN grant 10068.



\begin{thebibliography}{99}



\bibitem{novo}
K. S. Novoselov {\em et al.}, Nature {\bf 438}, 197 (2005).

\bibitem{zhang}
Y. Zhang {\em et al.}, Nature {\bf 438}, 201 (2005).

\bibitem{egger}
R. Egger and A. O. Gogolin, Phys. Rev. Lett. {\bf 79}, 5082
(1997).

\bibitem{kane}
C. Kane, L. Balents and M. P. A. Fisher, Phys. Rev. Lett. {\bf
79}, 5086 (1997).

\bibitem{mele1}
D. P. DiVincenzo and E. J. Mele, Phys. Rev. B {\bf 29}, 1685 (1984).

\bibitem{nos}
J. Gonz\'alez, F. Guinea and M. A. H. Vozmediano,
Nucl. Phys. B {\bf 406}, 771 (1993).

\bibitem{mele2}
C. L. Kane and E. J. Mele, Phys. Rev. Lett. {\bf 78}, 1932 (1997).

\bibitem{prl}
J. Gonz\'alez, F. Guinea and  M. A. H. Vozmediano,
Phys. Rev. Lett. {\bf 69}, 172 (1992).

\bibitem{graph1}
N. M. R. Peres, F. Guinea, and A. H. Castro Neto,
Phys. Rev. B {\bf 73}, 125411 (2006).

\bibitem{graph2}
V. P. Gusynin and S. G. Sharapov, Phys. Rev. Lett. {\bf 95},
146801 (2005).

\bibitem{lev}
D. S. Novikov and L. S. Levitov, Phys. Rev. Lett. {\bf 96},
036402 (2006).

\bibitem{dress}
R. Saito, G. Dresselhaus, M. S. Dresselhaus,
{\em Physical Properties of Carbon Nanotubes}, Chap. 6,
Imperial College Press, London (1998).


\bibitem{lu}
J. P. Lu, Phys. Rev. Lett. {\bf 74}, 1123 (1995).

\bibitem{add}
R. Saito, G. Dresselhaus, and M. S. Dresselhaus, Phys. Rev. B {\bf 53},
10408 (1996).

\bibitem{cuni}
N. Nemec and G. Cuniberti, Phys. Rev. B {\bf 74}, 165411 (2006).



\bibitem{ando}
H. Ajiki and T. Ando, J. Phys. Soc. Jpn. {\bf 62}, 1255 (1993);
J. Phys. Soc. Jpn. {\bf 65}, 505 (1996).

\bibitem{novikov}
H.-W. Lee and D. S. Novikov, Phys. Rev. B {\bf 68}, 155402 (2003).

\bibitem{mac}
J. W. McClure, Phys. Rev. {\bf 104}, 666 (1956).


\bibitem{brey}
L. Brey and H. A. Fertig, report cond-mat//0603107.


\bibitem{halp}
We connect here with the standard derivation of the
quantization of the Hall conductivity in a nonrelativistic
system, as given by
B. I. Halperin, Phys. Rev. B {\bf 25}, 2185 (1982).

\bibitem{ando2}
H. Akera and T. Ando, Phys. Rev. B {\bf 39}, 5508 (1989).

\bibitem{vk}
E. Ahlswede {\em et al.}, Physica B {\bf 298}, 562 (2001).

\bibitem{butt}
M. B\"uttiker, Phys. Rev. B {\bf 38}, 9375 (1988).



\end{thebibliography}
\end{document}